\documentclass{article}
\usepackage{amsmath}
\usepackage{amscd}
\usepackage{latexsym}
\usepackage{amssymb}

\usepackage{color}

\title{On the deformation of path algebras}
\author{Murray Gerstenhaber}

\begin{document}
\maketitle
\newtheorem{theorem}{Theorem}
\newtheorem{corollary}{Corollary}
\newtheorem{lemma}{Lemma}
\renewcommand{\abstractname}{}
\newcommand{\C}{\ensuremath{\mathbb{C}}}
\newcommand{\R}{\ensuremath{\mathbb{R}}}
\newcommand{\Z}{\ensuremath{\mathbb{Z}}}
\newcommand{\cP}{\ensuremath{\mathcal{P}}}
\newcommand{\cM}{\ensuremath{\mathcal{M}}}
\newcommand{\bM}{\ensuremath{\mathbf{M}}}
\newcommand{\cX}{\ensuremath{\mathcal{X}}}
\newcommand{\cA}{\ensuremath{\mathcal{A}}}
\newcommand{\bA}{\ensuremath{\mathbf{A}}}
\newcommand{\cB}{\ensuremath{\mathcal{B}}}
\newcommand{\pr}{\ensuremath{\preceq}}
\newcommand{\op}{\ensuremath{\mathrm{op}}}
\newcommand{\bfk}{\ensuremath{\mathbf{k}}}
\newcommand{\g}{\gamma}
\newcommand{\G}{\Gamma}
\newcommand{\z}{\zeta}
\newcommand{\Aut}{\operatorname{Aut}}
\newcommand{\dr}{\operatorname{dR}}
\vspace{-7mm}
\date{}
{}

\begin{abstract}\noindent Those elements of the second de Rham cohomology group of a  connected, oriented Riemannian manifold which map its second homotopy group to zero or to a discrete subgroup of the reals induce deformations of the path algebra of the manifold. If the image is not identically zero, then the induced deformations are quantized.  We examine the simplest examples, namely, the torus and the 2-sphere, and consider possible physical interpretations of the deformations of their path algebras.  
\end{abstract}

\section{Introduction} 

Throughout this paper $\cM$ will denote a connected, orientable Riemannian manifold.  Our main result is that those elements of the second de Rham cohomology group $H^2(\cM)$  which map  its second homotopy group $\pi_2$ to zero or to a discrete subgroup of the reals induce deformations of the path algebra $\cA(\cM)$ of $\cM$. Some induced deformations may be trivial.
The Hurewicz morphisms map the homotopy groups $\pi_n$ of $\cM$ into $H_n(\cM)$, and as  $H^n(\cM)$ is dual to $H_n(\cM)$, every  closed 2-form $\omega$ on $\cM$ also gives rise to a morphism $\pi_n \to \R$ which depends only on the cohomology class of $\omega$.  If $\omega(\pi_2) = 0$, then $\omega$ induces a family of deformations of $\cA$ which is continuous, while if $\omega(\pi_2)$ is not zero but is still a discrete subgroup of $\R$ then the family is quantized by the integers. The global structure of space may thus determine certain quantum phenomena. As the deformations induced by $\omega$ depend only on its cohomology class, here always denoted by $\bar\omega$,  the deformations may also be viewed as induced by $\bar\omega$.

The product of paths is their concatenation, or zero if they can not be concatenated. A deformation induced by an element $\bar\omega$ of $H^2(\cM)$ will multiply this concatenation product by a non-zero real or complex ``weight'' which is equal to one before deformation.  It generally remains equal to one if the two paths are geodesics and either their concatenation is still a geodesic, the second just being a continuation of the first, or if the second just reverses the direction of the first (as if a particle traveling along the first geodesic has at some point been reflected back on the path from which it came).  In the quantized case the weights have absolute value one and may be interpreted as phases; in the non-quantized case it seems natural to view them as real but in principle they can take on any real or complex value. 

Suppose that a particle or object has traveled a path $\g$ from $p$ to $q$ and now is traveling on a second path $\g'$ beginning at $q$, which we may suppose has been parameterized by its length from $q$. When the particle has covered a distance $x$ on $\g'$, arriving at a point $\g'(x)$, the weight attached to the product of $\g$ and the segment of $\g'$ from $q$ to $\g'(x)$ is a function of $x$ which depends on the path $\g$ that the particle initially traversed. The deformation has thus caused the particle to have some ``memory" of the first path it traversed;  this memory is a change in the particle. The deformation of path algebras suggests ways in which particles may change as they move.  

The results for the 2-sphere here apply also to  Euclidean 3-space with a point removed (representing, e.g., a body which the moving particle can not penetrate) or 4-space with a line removed (the space-time path of the body) and have obvious extensions to 3-space with any number of points removed. 

That the concept of deformation is essential to understanding quantization was first made explicit in the foundational paper \cite{BFFLS}; this note presents another aspect of the deformation approach to quantization.  

\section{Path algebras} By a \emph{path} on the Riemannian manifold $\cM$ we will mean a piecewise differentiable map of a directed segment of the real line into $\cM$.  The image then has a  well-defined length $\ell$, so parameterizing the image by its length from the starting point we have, in effect, a map $\g: [0,\ell] \to \cM, \ell \ge 0$ such that for all $0 \le x \le \ell$ the length of $\g([0,x])$ is exactly $x$. If $\ell = 0$ (which must be allowed) then the path is reduced to a point. Suppose that $\g: [0,\ell] \to \cM$ and $\g': [0,\ell'] \to \cM$ are two paths on $\cM$.  If $\g(\ell) = \g'(0)$ then we say that the paths can be \emph{concatenated} and we define their product $\g\g'$, also called their concatenation, by setting $\g\g'(x)= \g(x)$ for $0\le x \le \ell$ and $\g\g'(x) = \g'(x-\ell)$ for $\ell \le x \le \ell+\ell'$. The product is zero otherwise. 

With this multiplication, the paths on $\cM$ together with 0 form a monoid $\bM$, that is, a set with a single associative multiplication but which does not necessarily contain a unit element.  If we have a monoid $\bM$ and a commutative unital ring $\bfk$ then we can form the \emph{monoid algebra} $\bfk\bM$ whose underlying module is the free $\bfk$-module generated by the elements of $\bM$ (the set of all formal finite sums of elements of $\bM$ with coefficients in $\bfk$), with multiplication defined by that in $\bM$. When $\bM$ is the monoid of paths on $\cM$ and $\bfk = \R$ this is the \emph{path algebra} of $\cM$.  Those paths which are piecewise geodesic segments form a subalgebra. The  path algebra of $\cM$, here denoted $\cA(\cM)$, does not have a unit element. It would have to be the sum of all the points of $\cM$, which as elements of the path algebra are orthogonal idempotents, but one allows only finite sums. Classical path algebras, which have been studied in connection, e.g., with graphs and quivers, generally do have units. There is an extensive discussion of path algebras on the web.

An algebra $A$ over a field having a multiplicative basis, i.e., a basis $\cB$ such that the product of any two elements of $\cB$ is again in $\cB$ or zero, is in particular a monoid algebra. Matrix algebras, poset algebras, group algebras, and algebras of finite representation type over algebraically closed fields  are examples. (For the latter cf \cite{FiniteRep}.) 

Elements of the second de Rham cohomology group $H^2(\cM)$ will induce deformations of the path algebra of $\cM$.  One can introduce a Riemannian structure on any paracompact differentiable manifold and for the qualitative results that follow the choice of metric should not matter. In particular, if we have an element $\bar\omega \in H^2(\cM)$,  then the deformed path algebra it defines using one metric should, we conjecture, be isomorphic to that obtained with any other. For any numerical computations, however, one may need the metric.

\section{Cohomology of monoids} If we have  a monoid $\bM$  then its cohomology with coefficients in an abelian group $\G$, which for the moment we will write additively, is defined as follows. Let $\bM^n = \bM \times  \cdots\times \bM$ ($n$ times).  The $n$-cochains of $\bM$ with coefficients in $\G$ are mappings $F\!:\! \bM^n \to \G$; these form an additive group $C^n(\bM, \G)$. The coboundary operator $\delta: C^n \to C^{n+1}$ is defined by setting 
\begin{multline*}
\delta F(a_1,\dots,a_{n+1}) = F(a_2,\dots,a_{n+1})\, +\\ \sum_{i=1}^{n}(-1)^i F(a_1,\dots,a_{i-1}, a_ia_{i+1},a_{i+2},\dots,a_{n+1}) \\+ (-1)^{n+1}F(a_1,\dots,a_n).
\end{multline*}
Then $\delta\delta = 0$, the group $Z^n$ of $n$-cocycles is the kernel of $\delta$ on $C^n$, the subgroup of $n$-coboundaries $B^n$ is $\delta C^{n-1}$, and the $n$th cohomology group is $H^n(\bM, \G) = Z^n/B^n$. There are no 0-cochains. Unlike group cohomology, there is no operation of $\bM$ on the coefficient group $\G$. (One can not be consistently defined when there exist non-zero $a,b \in \cM$ with $ab = 0$.)

When $\G$ is multiplicative the coboundary formula can easily be rewritten in multiplicative form and the definitions remain the same. Suppose now that we have a monoid $\bM$ and coefficient ring $\bfk$, and that $f$ is a multiplicative 2-cocycle of $\bM$ with coefficients in the multiplicative group $\bfk^{\times}$ of units of $\bfk$. The cocycle condition can be rewritten in the form  $f(a,b)f(ab,c) = 
f(b,c)f(a,bc)$. (In this form one would not need that the values of $f$ be units.) With $f$ we can define a new multiplication on  the monoid algebra $\bfk\bM$ by setting  $a\!\ast\! b = f(a,b)ab$ for all $a, b \in \bM$ and then extending this bilinearly to all of $\bfk\bM$.  The cocycle condition insures that this multiplication is again associative.  We will call this a \emph{coherent deformation} of $\bfk\bM$. Note that $f$ can be multiplied by any element of $\bfk^{\times}$ so $f$ actually induces a ``one parameter'' 
family of coherent deformations parameterized by $\bfk^{\times}$. While these are not at first glance deformations in the classical sense of \cite{G:DefI} (cf also \cite{GS:Monster}), they will be shown to be closely related.  When $f$ is the coboundary of a 1-cochain, say $f =\delta g$, then the mapping of $\bfk\bM$ to itself sending $a \in \bM$ to $g(a)a$ is an isomorphism of $\bfk\bM$ with the $\ast$ multiplication to $\bfk\bM$ with its original multiplication. The deformation induced by $f$ is then called \emph{trivial}, but a trivial deformation may possibly still have physical significance.

An important special case of a coherent deformation is that where we start with an additive cocycle $F$ of $\bM$ with coefficients in $\R$.  Then we can define a multiplicative 2-cocycle $f$, the exponential of $F$, by setting $f(a,b) = \exp (F(a,b))$ and obtain thereby a family of coherent deformation of $\R\bM$. We may call $f(a,b)$ a \emph{weight} that has been put on the product $ab$. More generally, suppose that we have an additive 2-cocycle $F$ of $\bM$ with coefficients not in $\R$, but in $\R /\mu\R$ for some modulus $\mu$.  For every $n\in \Z$ we then have a well-defined multiplicative 2-cocycle $f$ defined by setting $f(a,b) = \exp((2n\pi i/\mu)F(a,b))$. The resulting family of coherent deformations is now quantized; the \emph{quantum number} of this $f$ is $n$.  In this case $|f(a,b)| = 1$ for all $a, b \in \bM$ and $f(a,b)$ may be interpreted as a phase.

To see the connection with classical algebraic deformation theory introduced in \cite{G:DefI} (as of this writing, exactly 50 years old), observe first that a cochain $\hat F$ in the Hochschild cochain complex $C^n(\bfk\bM,\bfk\bM)$ is completely determined by its values $\hat F(a_1,\dots, a_n)$ with $a_1,\dots,a_n \in \bM$, and conversely, giving those values will define a cochain.  So suppose that we have a monoid cochain $F\in C^n(\cM,\bfk)$. Then one can define an $n$-cochain $\hat F$ in  $C^n(\bfk\bM,\bfk\bM)$ by setting $\hat F(a_1,\dots, a_n)= F(a_1,\dots, a_n)a_1a_2\cdots, a_n$.  These cochains form a subcomplex of $C^n(\bfk\bM,\bfk\bM)$ and it is easy to check that we have a cochain mapping. Conversely, those Hochschild $n$-cochains $\hat F$ such that $\hat F(a_1,\dots, a_n)$ is just some multiple $\lambda (a_1a_2\dots a_n), \,\,\lambda \in \bfk$, of $a_1a_2\dots a_n$ whenever $a_1,a_2,\dots, a_n \in \bM$ form a subcomplex of the Hochschild cochain complex $C^n(\bfk\bM,\bfk\bM)$.
Sending $\hat F$ to $F \in C^n(\cM,\bfk)$ defined by setting $F(a_1,\dots,a_n) = \lambda$ is the inverse map. 

If $A$ is an arbitrary associative algebra then it was shown in \cite{G:DefI} that its second Hochschild cohomology group $H^2(A,A)$ with coefficients in itself is the set of infinitesimal deformations of $A$. It follows that the elements of $H^2(\bM,\bfk)$ can also be viewed as infinitesimal deformations of $\bfk\cM$ in the sense of \cite{G:DefI}. Recall now that it is often quite advantageous to compute the Hochschild cohomology of an algebra not from the full Hochschild cochain complex, but using a subcomplex defined by taking the cohomology relative to a separable subalgebra, cf. e.g. \cite{GS:Simplicial}.  Using this technique it is known that for finite poset algebras, which are in particular monoid algebras, the inclusion of the subcomplex of the Hochschild complex just defined into the full Hochschild complex is a quasiisomorphism, i.e., induces an isomorphism in cohomology. Because of the absence of a unit and the presence of infinitely many idempotents in the path algebra, there is no guarantee that here the inclusion of this subcomplex into the full Hochschild complex induces an isomorphism of  cohomology groups. We conjecture that it does, at least if we somehow take into account that $\bfk\bM$ carries a topology. If so, then $H^2(\bM, \bfk)$ would in fact be the full group of infinitesimal deformations of $\bfk\bM$ when the latter is naturally considered as a topological algebra.

\section{Modular exterior forms and deformations of the path algebra}  Let $\omega$ again denote a closed 2-form on $\cM$ and $\bar\omega \in H^2(\cM)$ be  its cohomology class. If $\bar\omega(\pi_2)$ is zero or a discrete subgroup of $\R$ then both $\omega$ and its class $\bar\omega$ will be called \emph{modular}. The \emph{modulus} of $\bar\omega$, denoted $\mu(\bar \omega)$, is then correspondingly defined to be zero or to be the least positive element of the image group, and is otherwise left undefined.  This definition can be extended to all dimensions. Those classes with rational moduli are dense in $H^n(\cM)$  and form a vector space over the rationals.

Returning to the case $n = 2$,  to every $\omega$ (or equivalently, to its class $\bar\omega$) we want to associate a 2-cochain, denoted $\tilde\omega$, of the path algebra $\cA$ of $\cM$.   It will only be necessary to define  $\tilde\omega(\g,\g')$ for pairs of paths $\g,\g'$ in $\cM$ since the paths form a multiplicative basis of the path algebra. However, this will not be possible for every pair of paths. Suppose that $\g$ is a path from $p$ to $q$ and $\g'$ a path from $q'$ to $r$.  If $q \ne q'$ then set $\tilde\omega(\g,\g') = 0$. If $q=q'$, then to make the definition we will need simultaneously that there is a unique shortest geodesic $\z$ homotopic to $\g$, that there is a unique shortest geodesic $\z'$ homotopic to $\g'$ and that there is a unique shortest geodesic $\z''$ homotopic to their concatenation $\g\g'$; if these conditions are not met, then $\tilde\omega(\g,\g')$ will be left undefined. 
To define $\tilde\omega(\g,\g')$ when the conditions are met, note that $\z''$ must also be homotopic to the concatenation of $\z$ and $\z'$. The homotopy then provides an element of area bounded by $\z, \z'$ and $\z''$. It is oriented by taking $p, q$ and $r$ in that order on the boundary, where $p, q$ are the starting and ending points, respectively, of $\g$, and $q, r$ are those of $\g'$. Then $\tilde\omega(\g,\g')$ is defined to be the integral of $\omega$ over this element. 
More precisely, the homotopy is a mapping from the unit square into $\cM$ and one can integrate the pull back of $\omega$ over the square.\footnote{The definition was misstated in v.1.}  

Note, however, that while the integral defining $\tilde\omega(\g,\g')$ is over a triangle whose sides are uniquely defined geodesics which depend only on the homotopy classes of $\g$ and $\g'$, the homotopy defining the element of area is not unique. If we have two distinct homotopies then they in effect define a mapping of $S^2$ into $\cM$. The difference between the integrals will be zero if this mapping is homotopic to zero but possibly not otherwise. There is, therefore, a fundamental condition that must be imposed on $\omega$, namely, that it have a modulus $\mu$. For then the difference between the integrals will be a multiple of the modulus $\mu$ of $\omega$ and the integral becomes well-defined and independent of the choice of homotopy if reduced modulo $\mu$.  Thus $\tilde\omega$ must be understood as having values in $\R/\mu\R$. (If $\mu = 0$ then the values are in $\R$.)

This $\tilde\omega$ will prove to be a cocycle. The fact that $\tilde\omega(\g,\g')$ may occasionally be undefined presents no serious problem if the set of cases in which it occurs is in some sense small. It is convenient,  in fact, to leave the product $\g\g'$ formally undefined if the conditions necessary to define $\tilde\omega(\g,\g')$ are not met, even though the concatenation is actually well-defined. Then $\tilde\omega(\g,\g')$ is defined precisely when the product $\g\g'$ is defined.  It  is natural to enlarge the concept of an algebra to allow that products be undefined in a small set of cases, and similarly for morphisms, cochains, and similar constructs. 
 (Since $\cM$ carries a volume form and hence a measure, so does $\cM \times \cM$. One`` smallness'' condition might be that the set of pairs of points $p,q$ such that some homotopy class of paths between them contains no unique shortest geodesic should have measure zero. We conjecture that this is in some sense almost always the case, if not always the case, but that might not be adequate. It can happen that $\cM$ has dimension 2 and that inside the 4-dimensional manifold $\cM \times \cM$ the set of such pairs of points $p,q$ has dimension 3, for example, the plane with a well. We want, at least, that it not disconnect $\cM \times \cM$, but what more may be needed is not known.)

Now  suppose that we have three paths $\g, \g', \g''$ such that the products $\g \g',\g' \g''$ and $\g \g'\g''$ are all defined, and consider $\tilde\omega$ for the moment as having values (as originally) in $\R$. The coboundary of $\tilde\omega$ evaluated on these three paths is 
\begin{equation*}
\delta \tilde\omega(\g,\g',\g'')=\tilde\omega(\g',\g'') -\tilde\omega(\g\g',\g'') +\tilde\omega(\g,\g'\g'')  -\tilde\omega(\g,\g').
\end{equation*}
This is just the integral of the closed 2-form $\omega$ over the surface of a 2-simplex in $\cM$ (which happens to have geodesic edges).   It is therefore a multiple of the modulus, so after reduction $\tilde\omega$  is in fact an additive 2-cocycle.  If $\mu = 0$ then introducing a parameter $\lambda$, the multiplicative 2-cocycle $\exp(\lambda\tilde\omega)$ gives a one-parameter family of coherent deformations of the path algebra $\cA(\cM)$ of $\cM$.   However, if  $\mu > 0$, then the multiplicative 2-cocycles $\exp((2n\pi i/\mu)\tilde\omega)$ define a discrete family of deformations indexed by $n\in \Z$; the deformations have been quantized. Note that since $\tilde\omega(\g,\g')$ depends only on the homotopy classes of $\g$ and $\g'$, all computations are actually taking place on the universal covering space of $\cM$. 

Some smaller algebras than the  path algebra should also be considered. The \emph{geodesic algebra} of $\cM$ is the free module generated by all directed geodesic segments on $\cM$, where the product is zero when two can not be concatenated and is otherwise the shortest geodesic homotopic to the concatenation when that geodesic is unique; otherwise it is undefined.  Here one sets $\tilde\omega(\g,\g')$ equal to the integral of  $\omega$ over the element of area defined by the homotopy, reduced, as before, by the modulus of $\omega$. The \emph{homotopy path algebra} of $\cM$ has as underlying module the free $\R$-module generated by triples $(p,q, [\g])$, where $(p,q)$ is an ordered pair of not necessarily distinct points of $\cM$ and $[\g]$ is a homotopy class of paths $\g$ from $p$ to $q$.  The product $(p,q,[\g])(q',r,[\g'])$ is zero if $q \ne q'$, and otherwise is $(p,r,[\g\g'])$. The path algebra maps onto the homotopy path algebra by sending every path to its homotopy class. The geodesic algebra is ``essentially'' isomorphic to the  homotopy path algebra, i.e., up to the omission of a ``small'' set, since it is just the homotopy path algebra with the multiplication left undefined when there is no unique shortest geodesic in $[\g\g']$. The definition of the homotopy path algebra is, like that of the path algebra, independent of the metric on $\cM$, but the set where the essential isomorphism is undefined does depend on the metric.  The homotopy path algebra has an obvious topology, as does the geodesic algebra; both map onto $\cM \times \cM$ and the map is locally a homeomorphism. However, there does not seem to be any obvious way to define $\tilde\omega$ for the homotopy path algebra.

It is not possible, in general, to define a morphism from all of $H^2(\cM)$ to the second monoid cohomology group of the path algebra $\cA(\cM)$ because of the varying moduli (or the lack of them) of the elements of $H^2(\cM)$. However, if we fix some non-negative $\mu \in \R$ and consider the additive subgroup of  $H^2(\cM)$ consisting of all elements whose modulus is an integral multiple of $\mu$, then there is a morphism of this subgroup into $H^2(\cA(\cM),\R/\mu\R)$. Its restriction to the monoid of paths can then be exponentiated, as above, to give a morphism into the second multiplicative cohomology group of that monoid with coefficients in $\R$ if $\mu = 0$, or in the circle group if $\mu > 0$.  This morphism is generally not a monomorphism and the deformation induced by a non-trivial element of $H^2(\cM)$ may, from the strictly algebraic point of view, be trivial, but as remarked might still have physical significance.  We don't know if the morphism is onto. If not, it would be interesting to know what other deformations there may be of the path algebra. 

The deformations defined here depend only on the class $\bar\omega \in H^2(\cM)$, but to compute the necessary integrals one must choose a representative form.  It seems natural, and most useful for computational purposes,  to choose the unique harmonic form which, by Hodge theory, is contained in the class.

As mentioned, the deformation theory of path algebras suggests ways in which a particle or body may change when it is moving. Suppose that we have a multiplicative 2-cocycle  $f$ of the path algebra and that a particle that has  traversed some fixed geodesic $\g$ from $p$ to $q$ is then deflected (e.g., by an observer) at the point $q$, continuing along a new geodesic $\g'$. Let the geodesic $\g'$ be parameterized by its length.
After traveling on $\g'$ for a distance $x$ the particle will have arrived at a point $\g'(x)$. Then the weight $f(\g,\g'(x))$ has become a function attached to the particle as a result of its travel.  While we have not defined the weight while the particle was on $\g$, but only once it is on the second geodesic $\g'$, note that $\g$ might be reduced to a point.  If $f$ has been obtained from a modular $\bar\omega \in H^2(\cM)$ with positive modulus then $|f(\g,\g'(\ell))| = 1$ for all $\ell$, so $f(\g,\g'(\ell))$ can be interpreted as a phase. The phase angle generally is not a linear function of $\ell$. It can vary discontinuously and it may happen that $f(\g,\g'(\ell))$ takes on only the values $\pm1$. This happens in the example of the 2-sphere.  In any case, the function $f(\g,\g'(\ell))$ preserves some memory of the original path $\g$.

The path on which a particle travels might also have some parameter other than its length, for example, the action in Legendre mechanics. If the probability that a particle will follow a particular path is dependent on how it arrived at the starting point of the path, then this dependency on how it got to the starting point  is again a change in the nature of the particle.

While we have explicitly shown only for dimension $n=2$ that modular elements of  $H^n(\cM)$ give rise to additive cocycles of the monoid of paths, which can then be exponentiated to give multiplicative ones, it is clear that the same is actually true in all dimensions. In dimension 2 these multiplicative cocycles have a natural interpretation as deformations of the path algebra. It would be useful to know how to interpret them in higher dimensions.

\section{Local triviality of deformations} The deformation induced by an element $\bar\omega \in H^2(\cM)$ is always trivial if one remains within the radius of injectivity of a point of $\cM$. (The radius of injectivity at a point $p \in \cM$ is the largest radius for which the exponential map at $p$ is a diffeomorphism. Equivalently, it is the distance to the cut locus, roughly, the set of other points to which there are multiple shortest geodesics.) If we have a path $\g$ from $p$ to $q$ which is shorter than the radius of injectivity at $p$, then there is a unique shortest geodesic $\z$ from $p$ to $q$.  It is necessarily homotopic to $\gamma$ so we can integrate $\omega$ over the element of area bounded by $\zeta$ and $\g$ to get a function $G(\g)$. Set $g(\g) = \exp(G(\g))$ if the modulus of $\omega$ is zero; if it is $\mu > 0$  then set $g(\g) = \exp((2\pi i/\mu)G(\g))$. The image of $\bar\omega$ in the group of multiplicative 2-cocycles of the monoid of paths is then just the coboundary of $g(\g)$. This exhibits the local triviality of the deformation induced by $\bar\omega$, but as $g(\g)$ may become undefined when one goes beyond the radius of injectivity, the deformation need not be trivial globally.

\section{The two-dimensional torus.}  In this example, the deformation induced by a non-zero $\bar\omega \in H^2(\cM)$ will be trivial since we will take the flat metric on the torus. We conjecture that this would still be the case more generally, a special case of the conjecture that the choice of metric doesn't really matter.  

Taking the flat metric on the 2-torus, it can be represented as the plane modulo a lattice. There is a unique geodesic in every homotopy class between any two points, which can be represented as straight line in the plane. In defining the path algebra one does not need to omit any pairs of points. The unique harmonic form $\omega$ is, up to constant multiple, just the usual element of area.  It is modular of modulus zero and there is a unique one-parameter family of deformations of the path algebra, given by $\exp(\lambda\tilde \omega)$, where $\lambda$ can take any real or complex value; there is no quantization of deformations.  

If $\g$ and $\g'$ are straight path segments which can be concatenated, and if the concatenation is again straight, so $\g'$ is either a continuation of $\g$ or a reflection going in the reverse direction, then $\tilde\omega(\g,\g') = 0$.  There is therefore no evidence of the deformation as long as a particle is moving in a straight line or is reflected.  With $\lambda$ real, suppose, however, that we have an ensemble of particles that have initially traveled along $\g$ and are now randomly deflected. The deformation may create a preference for one direction over another relative to the initial direction of travel. Since reflection of $\g'$ across the line defined by $\g$ replaces $\tilde\omega(\g,\g')$ by its negative, it also replaces the weight it defines by its inverse. The weight might conceivably be interpreted as the ratio of the probability of being deflected a fixed given distance in one direction from the initial direction of travel to the probability of the mirror image deflection.

The deformation, nevertheless, is algebraically trivial.  For recall, in the definition of the cocycle $\tilde\omega$ in the previous section, that when a ``shortest'' geodesic is chosen it is always the shortest one in a specific homotopy class so the construction is taking place, in effect, on the universal cover of $\cM$.  The de Rham cohomology of the plane vanishes, so the deformation is trivial.

The modulus here is zero (as it must be).  This, however, raises the question of whether an element of $H^2(\cM)$ with modulus zero always induces an algebraically trivial deformation.

\section{The unit sphere}

On the unit sphere $S^2$, geodesics between antipodal points are not unique so the product of concatenatable paths will be undefined when the beginning of the first is antipodal to the end of the second. However, the set of pairs of antipodal points in $S^2 \times S^2$ is small; it has codimension two since the set of antipodal points is an image of the sphere inside the product of the sphere with itself. The area form $\omega$ is, up to constant multiple, the only harmonic $2$-form. It is modular with modulus $4\pi$, the area of the sphere, so deformations of the path algebra here are quantized and given by the multiplicative cocycles $\exp((n/2)i\tilde\omega), n \in \Z$. 

Suppose now that we have a deformation with quantum number $n$. Consider a particle starting in the northern hemisphere of the unit sphere at a point at 0 degrees longitude which moves south on that meridian to the equator and then is deflected eastward, continuing to travel on the equator. As it continues to circle the equator it will experience a change in phase, the total change in angle on returning to the equatorial point at 0 degrees longitude at which it was deflected being $n\pi$. This change is not a linear function of the distance traveled along the equator unless the particle started at the north pole. The change is most rapid as the particle passes the point on the equator at longitude 180 degrees. It is, however, continuous, with one exception, that in which the initial path to the equator had length zero, i.e., if the particle started on the equator. The phase angle is then 0 until it reaches the antipodal point, when it becomes undefined. Thereafter, until it returns to the starting point, the phase angle is $n\pi$.  If the quantum number is even, then the angle remains constant at zero modulo $2\pi$.  If the quantum number is odd, however, then there is a sign that has been attached to the particle which is $+1$ from the start until the particle reaches the antipodal point, where it undefined, and then switches to $-1$ until the particle returns to the start.  

In the foregoing, the sign attached to a particle simply circling the equator switches exactly once in a full orbit, independent of the value of $n$, as long as $n$ is odd. The changes in sign occur even though the particle has not been deflected. By contrast, if the particle has not started at the equator but has been deflected there to travel along the equator, then the change in phase depends on $n$.  For even values of $n$, particles which are not deflected experience no change in phase, so the deformation is not apparent until they are deflected.  In either case, particles starting at a common point and ending at a common point may arrive with different phases, depending on their paths.

\section{Background} We give here briefly some background.  Suppose that we have a  compact manifold (not necessarily Riemannian) $\mathcal{X}$. By a theorem of J. H. C. Whitehead \cite{JHCWhite:Triang},  it then has a finite triangulation and the combinatorial structure is essentially unique.  Choosing one, we have a simplicial complex $K$ whose underlying space (or geometric realization) $|K|$ is homeomorphic to $\mathcal{X}$. The simplicial cohomology of $K$ is then isomorphic to the \v{C}ech cohomology of  $|K|$, which in turn is isomorphic to the de~Rham cohomology of $\mathcal{X}$.  Now the set of all simplices of $K$ becomes a poset $\cP$ under the relation $\sigma \pr \tau$ if $\sigma$ is a face of $\tau$.  Viewing $\cP$ as a category, denote its nerve by $\Sigma(\cP)$. Then $\Sigma(\cP)$ is just the barycentric subdivision of $K$, so its simplicial cohomology is again isomorphic to the de Rham cohomology of the manifold $\mathcal{X}$.  On the other hand, choosing any coefficient ring $\bfk$, we can form the poset  algebra $\bfk\cP$. By a theorem of the author and S. D. Schack (cf. \cite{GS:Simplicial}), the cohomology of this algebra with coefficients in itself is isomorphic to the simplicial cohomology of $\Sigma(\cP)$ with the coefficients taken in the same ring $k$.  Therefore, when $k = \R$, it is isomorphic to the de~Rham cohomology of $\mathcal{X}$. 

The algebra $\bfk\cP$ depended on the choice of the triangulation $K$, but if $K'$ is a subdivision and $P'$ the associated poset then there is an algebra morphism $\bfk\cP \to \bfk\cP'$.  This is not derived from a morphism of posets since simplices of $\cP$ are sent to sums of simplices of $\cP'$, but it does induce a morphism of Hochschild cochain complexes, cf. \cite{G:SelfDual}.  One would expect this to be a quasiisomorphism, since the complexes have the same cohomology, namely that of $K$.  Since by Whitehead's theorem any two triangulations have a common refinement, there is actually a direct limit to the algebras $\bfk\cP$ as well as to their Hochschild complexes. There is thus an algebra, denote it for the moment by $\bA$, intrinsically associated to the manifold $\mathcal{X}$.  We do not presently know its properties.  When $\mathcal{X}$ is the  2-sphere, the deformation theory of $\bA$ as an abstract algebra can not be identical with that of the path algebra as we have defined it, because the latter is quantized. However, the algebra $\bA$ should carry some topology and we do not yet know enough about the deformation theory of topological algebras. It is conceivable that the deformations of some topological algebras are necessarily quantized.  In any case, the existence of $\bA$ suggests looking for some algebra definable directly from the geometry of $\mathcal{X}$ whose deformations are governed by a cohomology also definable directly from the geometry of $\mathcal{X}$.  For orientable $\mathcal{X}$ the natural  choice seems to be the de Rham cohomology, but we do not yet have a natural definition of the algebra, if indeed it exists.
\medskip

The author thanks Jim Stasheff for helpful comments and suggestions.

\nocite{HazGer}

\end{document}